\documentclass[draft]{agujournal2019}
\usepackage{url} 
\usepackage{lineno}
\usepackage[inline]{trackchanges} 
\usepackage{soul}

\draftfalse

\journalname{JGR: Space Physics}

\begin{document}

\newpage

\title{The Structure of the warped Io Plasma Torus constrained by the Io Footprint}

\authors{Stephan Schlegel \affil{1}, Joachim Saur \affil{1}}

\affiliation{1}{Institut für Geophysik und Meteorologie, Universität zu Köln, Cologne, Germany}

\correspondingauthor{Stephan Schlegel}{sschleg1@uni-koeln.de}

\begin{keypoints}
\item	 Based on the Io Footprint positions, we show quantitatively that the Io Plasma Torus is centered around the centrifugal equator of Jupiter's multipole magnetic field
\item	Position of the Io Footprint can be used to constrain a density model of the Io Plasma Torus
\item  The displacement of the Io Plasma Torus due to higher magnetic field moments can change the plasma density at Io by up to 20\%
\end{keypoints}

\begin{abstract}
Standard models of force balance along Jovian field lines predict the location of the Io plasma torus to be the centrifugal equator of Jupiter's magnetosphere, i.e. the position along the magnetic field lines farthest away from Jupiter's rotational axis. In many models, the centrifugal equator is assumed to lay on a plane, calculated from a (shifted) dipole magnetic field, rather than on a warped surface which incorporates Jupiter's higher magnetic field moments. In this work, we use Hubble Space Telescope observations of the Io Main Footprint to constrain density, scale height and lateral position of the Io Plasma Torus. Therefore, we employ the leading angle of the footprints to calculate expected travel times of Alfv\'en waves  and carry out an inversion of the observations. For the magnetic field we use the JRM33 magnetic field model. The inversion results show peak densities between $\rho_0 = 1830$~cm$^{-3}$ and $\rho_0 = 2032$~cm$^{-3}$ and scale heights between $H = 0.92 R_J$ and $H = 0.97 R_J$ consistent with current literature values. Using a warped multipole centrifugal equator instead of a planar dipole increases the quality of the fit by about $25 \%$. We additionally develop two tests to confirm that the multipole centrifugal equator from the JRM33 model fits explains the applied data set better than the dipole centrifugal equator.  The quadropole moments alter Io's relative position to the torus, which changes the plasma density around Io by up to $\Delta \rho / \rho = 20\%$.

\end{abstract}

\section{Introduction}

Io's interaction with the surrounding plasma is an important feature of Jupiter's inner magnetosphere. On the one hand it feeds the Io Plasma Torus by atmospheric sputtering (e.g. \citeA{haff1981sputter,mcgrath1987magnetospheric,saur2004plasma,bagenal2020space} and references therein), where ion-neutral collisions eject particles from Io's atmosphere that generate a neutral torus in Io's orbit. This neutral torus gets successively ionized, forming the Io Plasma Torus. Furthermore, the plasma locally around Io is perturbed by the collision with Io and its atmosphere. These perturbations travel as Alfv\'en waves along the magnetic field lines and accelerate particles close to Jupiter's ionosphere \cite{crary1997generation, damiano2019kinetic, szalay2018situ, szalay2020new, janser2022properties}. The accelerated particles travel along the magnetic field lines, generating aurora at both hemispheres \cite{hess2010power, bonfond2015far, saur2013magnetic, schlegel2022alternating}, called the Io Footprint. The location of these footprints depends on the magnetic field model and density model along the magnetic field line and have been used to constrain the VIP4 magnetic field model \cite{connerney1998new}. With the in-situ magnetic field measurements from the Juno spacecraft, a precise magnetic field model for the inner Jovian magnetosphere up to $30^{th}$ degree is available now in the form of the JRM33 \cite{connerney2022new}.  Therefore, the position of the Io Footprint can now be used to constrain the density profile along the magnetic field lines and give insight about the density structure and location of the Io Plasma Torus. \\
The torus is often considered to lie at the centrifugal equator, the position along the magnetic field line farthest away from the rotational axis \cite{khurana2004configuration,thomas2004io}. In the case of a dipolar magnetic field, the centrifugal equator is planar, roughly $2/3$ on the way from the rotational equator to the magnetic equator. However, higher order moments warp the centrifugal equator "like a potato chip" \cite{phipps2020io, herbert2008new}. Other previous observation also show a more complex structure of the torus, not consistent with a dipole centrifugal equator \cite{bagenal1994empirical, schneider1995structure, phipps2020io}. However, the previous work did not demonstrate with quantitative measures that the torus is located at the multipole centrifugal equator. \\
The aim of this work is to quantitatively demonstrate that the plasma torus is centered around the multipole centrifugal equator. Therefore, we use the positions of the Io Footprint to constrain a density model of the Io Plasma Torus and its location depending on System III longitude. For that, we map Alfvén waves along the magnetic field lines and compare the resulting expected location of the footprint to Hubble Space Telescope observations and infer Alfv\'en wave travel times. We use these travel times as an input for an inversion and analyze the output regarding the hypothesis of a dipole or multipole centrifugal equator.


\section{Model and Methodology of the Inversion}\label{sec:Inversion}

\subsection{Location of the Io footprints}

When Jupiter's co-rotating plasma collides with Io and its tenuous atmosphere it gets perturbed. These perturbations propagate as Alfv\'en waves along the magnetic field lines that are frozen into the plasma. Close to Jupiter in the acceleration region, these waves cause wave-particle interaction and accelerate particles towards and away from Jupiter. The accelerated particles collide with molecules in Jupiter's upper atmosphere and create auroral emissions. Since the accelerated particles travel along the magnetic field lines and the Alfv\'en velocity close to Jupiter approaches the speed of light, the exact height of the acceleration region or the emissions does not affect the travel time significantly and we can assume that the emissions are created at the location where the Alfv\'en waves connect to Jupiter's atmosphere. Therefore, we assume that Io's main footprint is located at the position of Io's main Alfv\'en wing (MAW) on Jupiter's 1 bar level. \\
Since the Alfv\'en waves get reflected at phase velocity gradients, which are most prominent at Jupiter's ionosphere and the Io torus boundary, there is a multitude of secondary footprints. Furthermore, the particles in the acceleration region are also accelerated away from Jupiter, creating footprints on the opposing hemisphere, which can results in leading spots that are upstream from the MAW-footprint. This work only focuses on the location of the MAW-footprints, since the secondary spots are dependent on the reflection pattern and the leading spot is affected by broadening due to electron drifting of about $\Delta \varphi \approx 0.7^\circ$ corresponding to $\Delta l \approx 200$~km broadening of the leading spot on Jupiter's surface for high energy electrons with energies of $E_e = 1$~MeV \cite{mauk1997energy}. This results in a difficult determination of the exact position of the leading spot and its corresponding magnetic field line. \\
The location of the MAW-footprint can be calculated with the Alfv\'en characteristic

\begin{equation}
    z^\pm = \mathbf{v} \pm \mathbf{v}_A,
\end{equation}

with the plasma velocity $\mathbf{v}$ and the Alfv\'en phase velocity

\begin{equation}
    \mathbf{v}_A = \frac{\mathbf{B}}{\sqrt{\rho \mu_0}},
\end{equation}

depending on the magnetic field strength $\mathbf{B}$ and the plasma mass density $\rho$. Since at high latitudes, the plasma is very dilute a relativistic correction for the Alfv\'en velocity has to be implemented:

\begin{equation}
    v_A^* = \frac{v_A}{\sqrt{1 + v_A^2 / c^2}}
\end{equation}


\subsubsection{HST observations}

\begin{figure}
\begin{center}
\includegraphics[trim = 1cm 10.5cm 0cm 10.5cm, clip = true ,scale = 0.8]{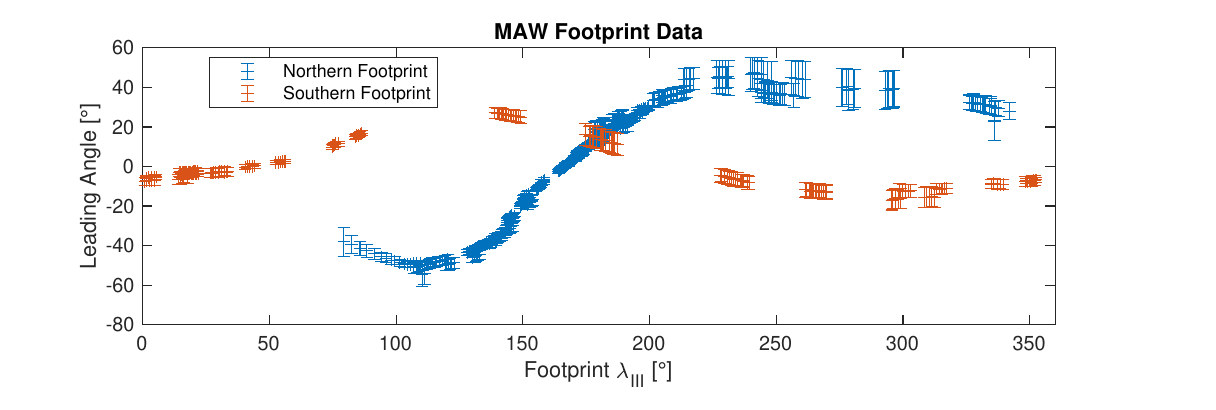}
\caption{Leading angles of the northern (blue) and southern (red) footprint, calculated from the observations published in \citeA{bonfond2017tails}. Many of the data points are clustered, especially visible for the southern footprint. The lack of observations between $0^\circ$ and $70^\circ$ for the northern footprint is because of the high angular velocity of the footprint in this area. Therefore, the footprint remains at this range only for a short time ($\approx 50$~min).}
\label{fig:Data}
\end{center}
\end{figure}

The position of the footprints relative to Io can be described as leading angle $\varphi = \varphi_{Io} - \varphi_F$, which is the longitudinal difference between Io's orbital position $\varphi_{Io}$ and the Io footprint $\varphi_{F}$ in System-3 coordinates. The positions here are projected to a height of 900~km above the 1~bar level of Jupiter. The data used here has been published as supplementary material by \citeA{bonfond2017tails} and is shown in Figure \ref{fig:Data}. The observations have been mostly conducted between February and June 2007. For the northern footprint additional data from 2005 and 2006 has been used. The errors $\varepsilon_{\varphi}$ are mostly due to inaccuracies in the determination of Jupiter's position using the limb fitting method as described in \citeA{bonfond2009io}. This likely leads to systematic errors in the longitudinal position of the footprints. Furthermore, the observations of the same visit can not necessarily be regarded as independent from each other. This would mean that the errors of clustered data might be correlated. Close to the limb of Jupiter, the error bars grow larger on account of projection effects. 

\subsubsection{The magnetic field model}

\begin{figure}
\begin{center}
\includegraphics[trim = 2cm 10cm 2cm 10cm, clip = true ,scale = 1]{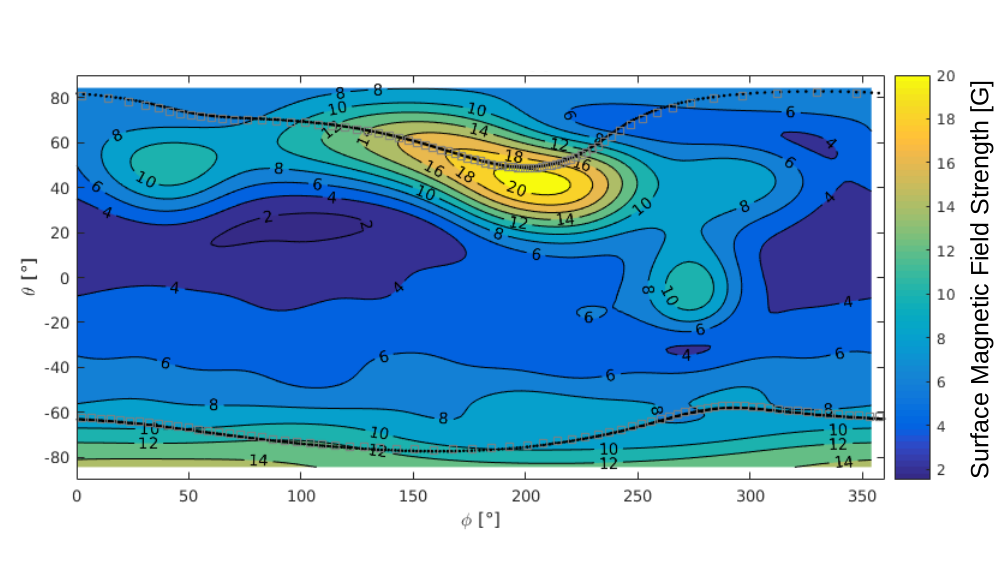}
\caption{The magnetic field strength on the flattened surface of Jupiter, calculated with the JRM33 model \cite{connerney2022new}. The black dots indicate the trajectory of the Io footprint in the northern and southern hemisphere in $1^\circ$ longitudinal separations along Io's orbit. The grey squares are the observational positions of the Io main footprints.}
\label{fig:JRM33}
\end{center}
\end{figure}

The Alfv\'en waves travel along the magnetic field lines that in this model are assumed to be fixed in Jupiter's rotating frame. Therefore, the location of the footprints only depend on the magnetic field lines connecting Jupiter's ionosphere to Io's orbit. This leads to all Io footprints to be confined to one line on the surface of each Jovian hemisphere. Though the magnetic field in Io's vicinity can often be regarded as a dipole of strength $M = 4.177$~G and a latitudinal tilt of $\vartheta_D = 10.25^\circ$ in $\varphi_D = 196.38^\circ$ western longitude, the magnetic field closer to Jupiter is more complex. We calculated the footprint trajectories as shown as black lines in Figure \ref{fig:JRM33} using the JRM33 magnetic field model by \citeA{connerney2022new}. This model has been created using the magnetic field data of the first 33 Juno flybys. we used all available Gauss-coefficients $g_l^m$ and $h_l^m$ up to degree $l = 30$ to map Io's orbit  to the dynamically flattened (1/15.4) surface  of Jupiter along the magnetic field lines to Jupiter's 1 bar level \cite{connerney2022new}.

\begin{figure}
\begin{center}
\includegraphics[trim = 2cm 10cm 2cm 10cm, clip = true ,scale = 1]{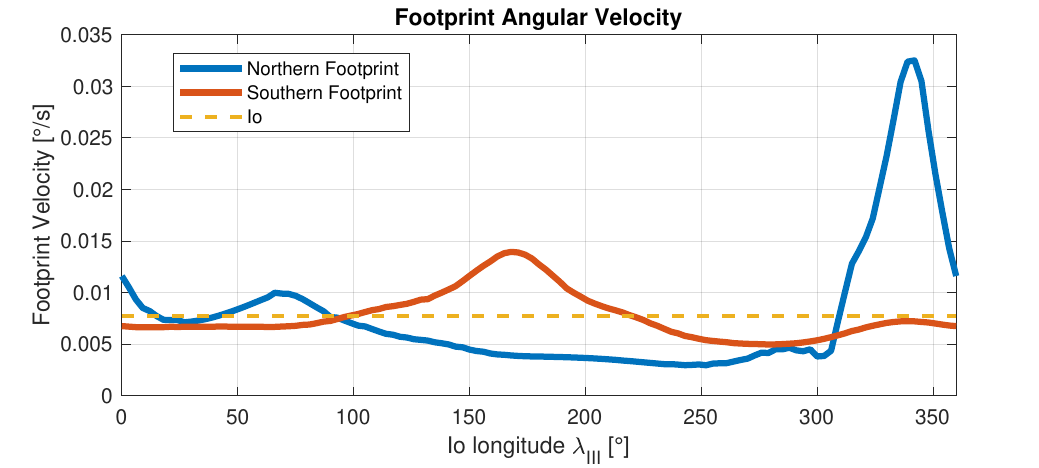}
\caption{The longitudinal angular velocity of the northern (blue) and southern (red) footprint. The synodical angular velocity of Io is shown as a reference as yellow dashed line at about $0.0077^\circ$/s. The northern magnetic field is more structered leading to a more variable angular velocity of the northern footprint. }
\label{fig:FPVelocity}
\end{center}
\end{figure}

\begin{figure}
\begin{center}
\includegraphics[trim = 4cm 8cm 4cm 7.4cm, clip = true, width = 0.8\textwidth]{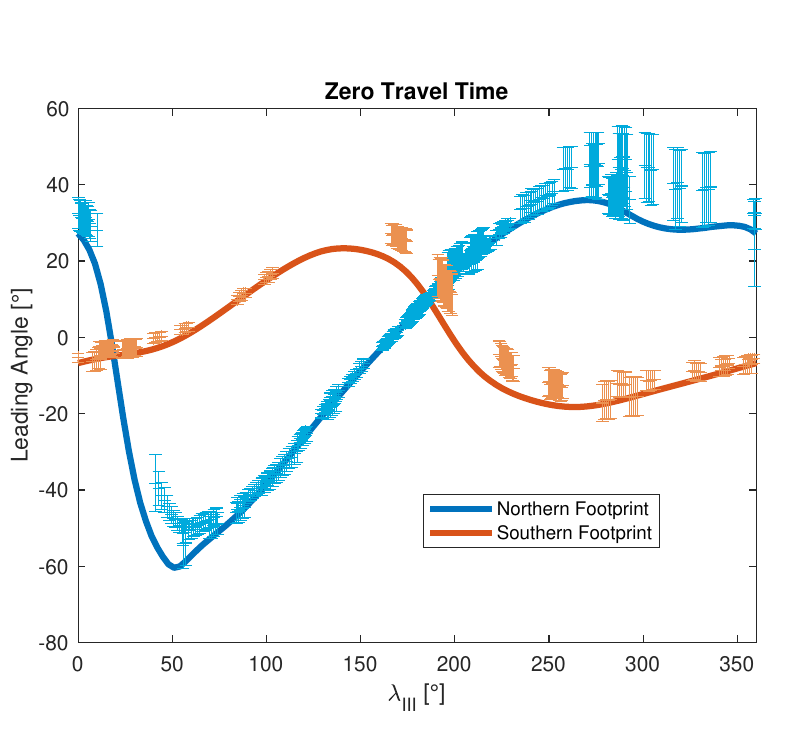}
\caption{The Leading Angle without travel travel time assumed for the northern (red) and southern (blue) footprint. The leading angle mostly underestimates the data (black with error bars), since the travel time increases the leading angle. This is especially apparent between 270 and 90$^\circ$ for the northern and 150 and 270$^\circ$ for the southern footprint, where Io should be closer to the southern and northern torus boundary, respectively.}
\label{fig:LA_mag}
\end{center}
\end{figure}

As can be seen, the footprints are generally drawn towards higher magnetic field strength. Since the magnetic field in the northern hemisphere is more complex than in the southern hemisphere, the trajectory there spans over a broader range of latitude ($45^\circ < \vartheta_{F}<83^\circ$). Furthermore, the separation between the footprint mappings is smaller where the magnetic field is stronger, which implies a slower movement of the Io footprint over Jupiter's surface as shown in Figure \ref{fig:FPVelocity}. There, the travel time has a lower influence on the leading angle $\varphi$ than at locations where the spacing is larger.  The leading angles $\varphi_B$ that only result from the magnetic field model are shown in Figure \ref{fig:LA_mag}, where no travel time of the Alfvén waves are assumed. Here, the change of the leading angles $\dot{\varphi}_B = \dot{\varphi}_{Io} - \dot{\varphi}_{F}$ only depends on the difference of the angular velocities of the footprints $\dot{\varphi}_F$ (solid lines in Figure \ref{fig:FPVelocity}) and Io $\dot{\varphi_{Io}}$ (yellow dashed line in Figure \ref{fig:FPVelocity}). Qualitatively, the observations (black with error bars) match the behaviour of the results of the calculations (solid lines). Since no travel time is included here, the calculations are generally underestimating the leading angles. Where the travel time has low influence, e.g. between 150$^\circ$ and $200^\circ$ for the northern footprint (blue), the observations are fairly well matched already. On the other hand, where travel time has a strong influence, e.g. close to $0^\circ$ for the northern footprint, the mapping strongly overestimates the observations.

\subsubsection{Influence of the Io Plasma Torus mass density}

The Io plasma torus is generally assumed to be centered around the centrifugal equator of Jupiter's magnetosphere, i.e. the position along the magnetic field lines that map towards Io's orbit and is the farthest away from Jupiter's rotation axis. A tilted or an offset tilted dipole results in the torus to be confined on a plane tilted by $\theta_C = 6.83^\circ$  in the direction of  $\varphi_D = 196.28^\circ$ western longitude. However, moments of higher degree, especially quadropole, still have an influence of the magnetic field at Io's orbit \cite{phipps2020io}. The discrepancy between of the latitudinal position of the centrifugal equator using the JRM33 full magnetic field model and only the dipole components can be up to $1.5^\circ$, which translates to about $\approx 0.15 R_J$ or $\approx 6 R_{Io}$. The centrifugal equator is a good estimate for the position of the plasma torus as it is derived from the force balance between pressure force and centrifugal force  along the magnetic field lines . \\
The torus itself is often regarded to be split in three parts (e.g. \citeA{bagenal2020space,phipps2018distribution} and references therein). The cold torus inside the orbit of Io, the ribbon region, where the plasma density is highest and warm torus starting roughly at the orbit of Io and decreases in density outwards. Io itself mostly is located inside the warm torus, but due to a dawn-dusk asymmetry, Io's orbit can cross into the ribbon region \cite{barbosa1983dawn}. \\
 The most widely used model for the density distribution $\rho$ is in the form of 
\begin{equation}\label{eq:densityModel}
    \rho(s) = \rho_0 \exp[-s^2/H^2],
\end{equation}
with a peak density $\rho_0$ at the centrifugal equator and a Gaussian decrease  with distance $s$ to the torus center along the magnetic field line \cite{gledhill1967magnetosphere, phipps2018distribution, phipps2021two, bagenal1994empirical}. This coincides with a force balance between centrifugal force and pressure gradient for an isothermal plasma. The scale height $H$ and plasma temperature $T$  are related \cite{thomas2004io} and can be   approximated by

\begin{equation}\label{eq:scaleHeight}
    H = \sqrt{\frac{2 k_B T}{3 \Omega_{J} \langle m \rangle}},
\end{equation}

with Jupiter's rotational frequency $\Omega_J$ and the mean ion mass $\langle m \rangle$. \citeA{dougherty2017survey} also use pressure anistropy, ambivalent electric fields and multiple species to derive a density distribution along the magnetic field line. 
However, in this work we will use a simplified density model of the form of Equation (\ref{eq:densityModel})  in order to reduce the amount of fitting parameters for the inversion. \\
\citeA{hinton2019alfven} used the JRM09 magnetic field model \cite{connerney2018new} and the CAN model \cite{connerney1981modeling} together with the density model by \citeA{dougherty2017survey} to calculate travel times from Io's orbit towards Jupiter. The authors fitted the travel times with a third degree Fourier series corresponding to 
\begin{equation}\label{eq:timeFit}
    t_{Fit}(\lambda_{III}) = \underbrace{A_0 + A_1 \cos(\lambda_{III} + a_1)}_{1^{st}} + \underbrace{A_2 \cos(2 \lambda_{III} + a_2)}_{2^{nd}} + \underbrace{A_3 \cos(3 \lambda_{III} + a_3)}_{3^{rd}}
\end{equation}
and found average travel times of 433~s and 401~s for the northern and southern hemisphere, respectively. The difference is due to the asymmetry of the magnetic field. 
In this work, we use the travel times to constrain a density model  corresponding to Equation (\ref{eq:densityModel}) with a peak density at the centrifugal equator.  To visualize the data  shown in Figure \ref{fig:Data} for that purpose more clearly, the leading angles have been converted to travel times $t_0$ using the synodic angular velocity $\Omega_{syn}$ of Io around Jupiter with

\begin{equation}\label{eq:travelTime}
    t_0 = \frac{\varphi_{Io} - \varphi_{F}}{\Omega_{syn}}.
\end{equation}

\begin{figure}
\begin{center}
\includegraphics[trim = 1cm 9.5cm 1cm 9.5cm, clip = true ,scale = 0.8]{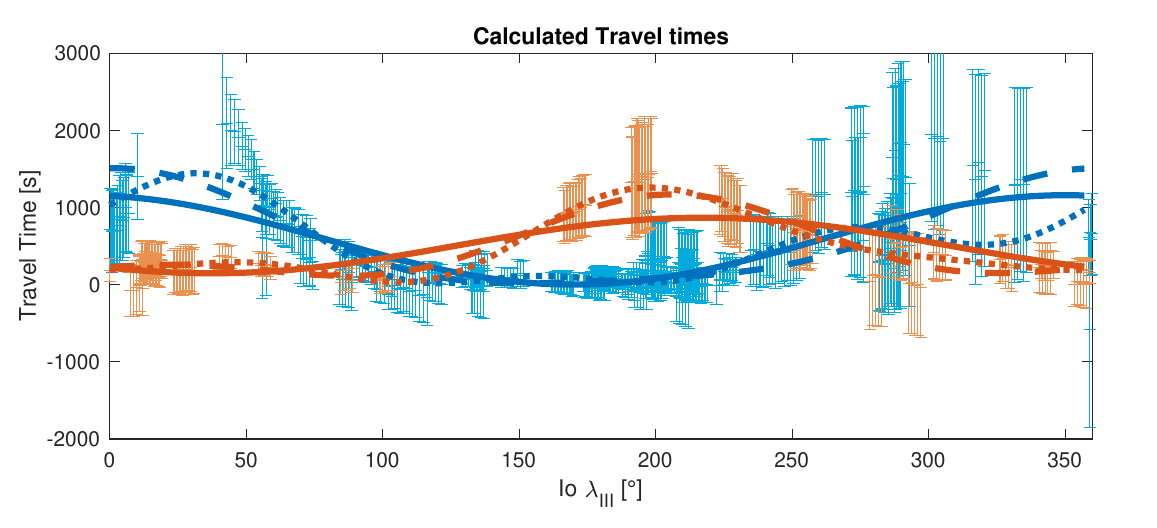}
\caption{The calculated travel times for the northern (blue) and southern (red) footprints with their corresponding error bar $\varepsilon_t$. The solid line is a first degree fit and the dashed lines is a second degree Fourier fit using Equation (\ref{eq:timeFit}).  The values are computed from the JRM33 \cite{connerney2022new} mapping and the footprint data published by \cite{bonfond2017tails}.}
\label{fig:TravelTimeFit}
\end{center}
\end{figure}

Furthermore, the errors are due to inaccuracies in the determination of the footprint positions, but not the position of Io. Therefore, the error in travel time $\varepsilon_t = \varepsilon_\varphi / \dot{\varphi}_F$ has to be weighted corresponding to the current longitudinal velocity of the footprint according to Figure \ref{fig:FPVelocity}. The calculated travel time data are depicted in Figure \ref{fig:TravelTimeFit}. The data has been fitted using a Fourier fit up to degree three corresponding to Equation (\ref{eq:timeFit}). The misfits $\chi = \sqrt{1/N \sum (t_0 - t_{Fit})^2 / \varepsilon_t^2}$ are 0.76, 0.68 and 0.65 for the northern and 0.63, 0.38 and 0.30 for the southern footprint for the fits of degree one, two and three, respectively. The fitting values are shown in Table \ref{tab:fitting} together with the values calculated from the model of \citeA{dougherty2017survey} by \citeA{hinton2019alfven}. 

\begin{table}
\caption{Fits of the travel time up to third degree according to Equation (\ref{eq:timeFit}), corresponding to the curves shown in Figure \ref{fig:TravelTimeFit}. As a reference, the third degree fit of the travel times calculated by \citeA{hinton2019alfven}, based on the model by \citeA{dougherty2017survey} are given.}\label{tab:fitting}
\centering
\begin{tabular}{l c c c c c c c}
\hline
Fit / Model & $A_0$ [s]& $A_1$ [s]& $a_1$ [$^\circ$]& $A_2$ [s]& $a_2$ [$^\circ$] & $A_3$ [s] & $a_3$ [$^\circ$] \\
\hline
First Degree North & 579.4 & -579.3 & -170.44 & 0 & 0 & 0 & 0 \\
Second Degree North & 603.6 & -728.2 & -178.3 & 175.0 & -12.3 & 0 & 0\\
Third Degree North & 534.5 & -526.6 &  178.9 & 246.0 &  2.8 &  23.0 & -100.5 \\
Hinton et al. North & 432.9 & 289.3 & -104.3 & 21.2 & 77.0 & 8.4 & 46.7 \\
\hline
First Degree South & 507.1 & 360.7 & 142.5 & 0 & 0 & 0 & 0\\
Second Degree South & 478.1 & 456.4 & 161.8 & 236.0 & -51.0 & 0 & 0\\
Third Degree South &  479.2 & 440.1 & 146.8 & 266.5 & -45.4 & 11.3 & -162.4\\
Hinton et al. South & 400.7 & 260.7 & 65.2 & 19.4 & -87.9 & 10.5 & -155.9\\
\hline
\end{tabular}
\end{table}

Overall the average travel time calculated from the footprint positions is slightly higher and the travel times are more variable compared  the values calculated from the model by \citeA{dougherty2017survey} .  The higher travel times indicate slower Alfvén velocities and therefore an overall higher plasma content of the torus. The higher variability of the travel times imply a larger influence of Io's relative distance to the torus center, which could either be explained by a more variable torus position or smaller scale height.  Another interesting fact is that the southern travel times are generally shorter and overall less variable due to the more homogeneous magnetic field in the South. Therefore, the southern travel times reflect the plasma density along the field line better than the northern travel times.  The variation of travel times can mostly be explained by a relative shift of Io's position with respect to the torus center. Therefore,  the strong decrease in misfit from first to second degree Fourier series already shows that a warped centrifugal equator due to quadropole moments fit the data much better than an offset dipole centrifugal equator. The fit of the northern footprints are mostly constrained by the observational data between 130$^\circ$ and 200$^\circ$ which has fairly small error bars. However, all fits show an $a_1$ value of around 180$^\circ$, which indicates that the torus is tilted in line with the dipole tilt of the JRM33 model of $\varphi_D = 196.38^\circ$. The fairly small decrease in misfit from second to third degree fits (0.03 for the northern and 0.08 for the southern footprints) hints that the position of the torus is mostly constrained by dipole and quadropole moments.


\subsection{Cost function and inversion method}

The travel time data, converted according to Equation (\ref{eq:travelTime}), is now fitted using a density model corresponding to Equation (\ref{eq:densityModel}). The cost function  $\Phi$ of this inversion scheme can be written as

\begin{equation}
    \Phi = \sum\limits_i \left( \frac{t_{0,i} - t_{\rho,i}}{\varepsilon_{t,i}} \right)^2,
\end{equation}
with the calculated travel times 
\begin{equation}
    t_\rho(\rho_0, H) = \int\limits_{Io}^{J} \frac{1}{v_A^*} ds
\end{equation}
mapped along the magnetic field line. It is important to note that the field line connected to the footprint is used since this is the field line that the Alfv\'en waves propagate on starting from Io's position. To minimize the cost function, a Monte-Carlo inversion method has been used to sweep the parameter space. For the scale height $H$ values between $H_{min} = 0.4 R_J$ and $H_{max} = 1.6 R_J$ and for the peak number density $n_0 = \rho_0 / \langle m \rangle$ values between $n_{min} = 500$~cm$^{-3}$ and $n_{max} = 3500$~cm$^{-3}$ have been used. With this approach, the sensitivity of the inversion towards the fitting parameters $H$ and $\rho$ as well as the correlation between them can be analyzed.


\section{Inversion Results}

In a first step the travel times are fitted for the peak density located at both, the dipole and the JRM33 multipole centrifugal equator and compared to values in the literature. In a second step, the position of the torus is fitted separately in another inversion to evaluate, whether the dipole or multipole centrifugal equator explains the data better.

\subsection{Best fit models}

\begin{figure}
\begin{center}
\includegraphics[trim = 1.5cm 9.5cm 2cm 10cm, clip = true , width = 1\textwidth]{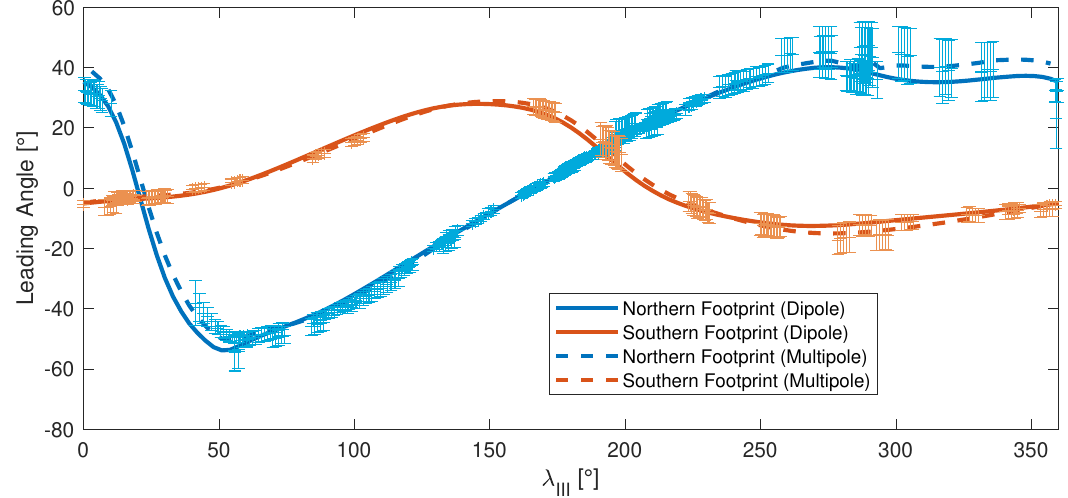}
\caption{The best fit models for the northern (blue) and southern (red) leading angles for both, the dipole (solid line) and multipole (dashed) centrifugal equator model. The multipole model generally fits the data better.}
\label{fig:BestFitBoth}
\end{center}
\end{figure}

\begin{figure}
\begin{center}
\includegraphics[trim = 1.5cm 9cm 1.8cm 9cm, clip = true , width = 1\textwidth]{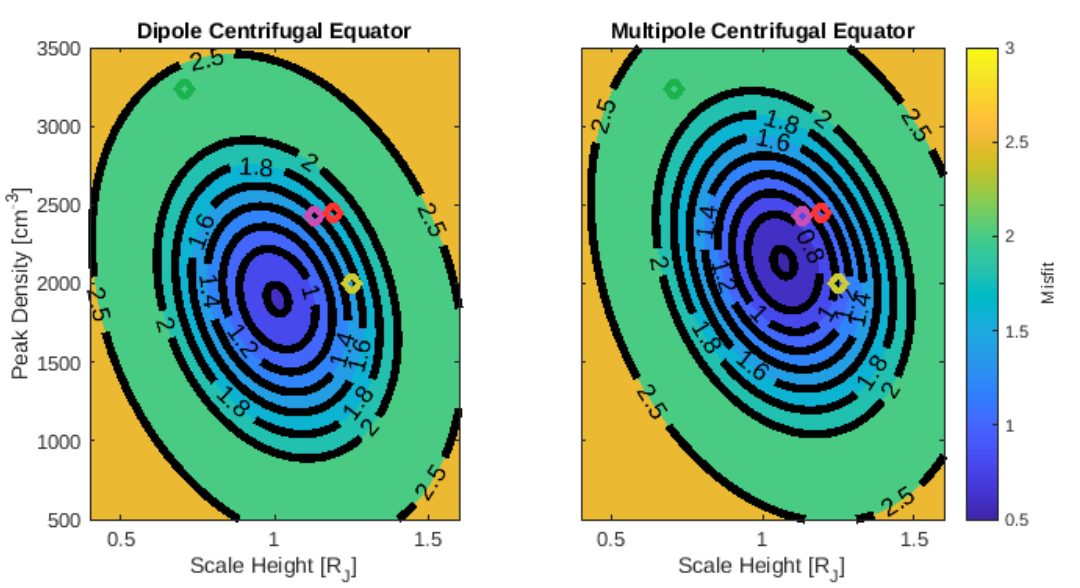}
\caption{Misfit contour of the Monte Carlo inversion for the dipole (left) and multipole (right) centrifugal equator model. The peak density and scale height for the warm torus (purple) and ribbon (green) of the model of \citeA{phipps2018distribution} as well the model by \citeA{bagenal1994empirical} (yellow) and \citeA{dougherty2017survey} (red) are indicated as diamonds. The scale height of the latter two are calculated with Equation (\ref{eq:scaleHeight}).}
\label{fig:MonteCarlo}
\end{center}
\end{figure}

For the first inversion the peak density $n_0$ is located at the JRM33 dipole and multipole centrifugal equator. The resulting leading angles are shown in Figure \ref{fig:BestFitBoth}. For the dipole model, the values for peak density and scale height are $n_0 = 1900$~cm$^{-3}$ and $H = 1.01 R_J$, while for the multipole model the values are $n_0 = 2133$~cm$^{-3}$ and $H = 1.07 R_J$, respectively. The two models do not differ much in travel times and therefore in leading angle. However, the misfit of $\chi = 0.58$ of the multipole best fit is considerably improved compared to the misfit of $\chi = 0.78$ of the dipole model. This is mostly due to some very low error observations of the southern footprint between 50$^\circ$ and 100$^\circ$ eastern longitude. For the southern footprint, the density model has a more consistent impact on the travel time and leading angle due to the longitudinal more homogeneous magnetic field. In Figure \ref{fig:MonteCarlo} the misfit for the whole Monte-Carlo inversion parameter domain is shown for both models. Since the errors of the observation are  considerably large and comparable to the overall travel time (compare Figure \ref{fig:TravelTimeFit}), a large parameter space can fit the observations with a misfit of $\chi < 1$. This allows us to estimate an uncertainty to the best fit parameters. For the dipole model, we find $\Delta n_0 = 321$~cm$^{-3}$ and $\Delta H = 0.13 R_J$. For the multipole model, the uncertainties are larger due to the overall better fit, and we get $\Delta n_0 = 413$~cm$^{-3}$ and $\Delta H = 0.17 R_J$. We can further compare the best fit models to the values given by \citeA{phipps2018distribution} for the warm torus and ribbon region and \citeA{dougherty2017survey} and \citeA{bagenal1994empirical} for the vicinity of Io's orbit, shown as stars in Figure \ref{fig:MonteCarlo}. Generally the values in the literature are higher in both peak density and scale height, but are mostly inside the $\chi < 1$ region for the multipole centrifugal equator. The results of the inversion overall show a good agreement with the literature, especially the results of the multipole model inversion.

To quantify the improvement of the multipole centrifugal equator, a Monte-Carlo Test was performed. In this test, each data point has been randomized with Gaussian noise  corresponding to their calculated error  added to their value. The number of data points that are fitted by one model rather than the other has been counted. This procedure has been repeated $N = 100000$ times. In the end, $91.4\%$ of randomized data points are fitted better by the multipole centrifugal equator and only $8.6\%$ of data points are fitted better by the dipole centrifugal equator model. 


\subsection{Position of the Io Plasma Torus}

\begin{figure}
\begin{center}
\includegraphics[trim = 1cm 5.5cm 0cm 5.5cm, clip = true , width = 1\textwidth]{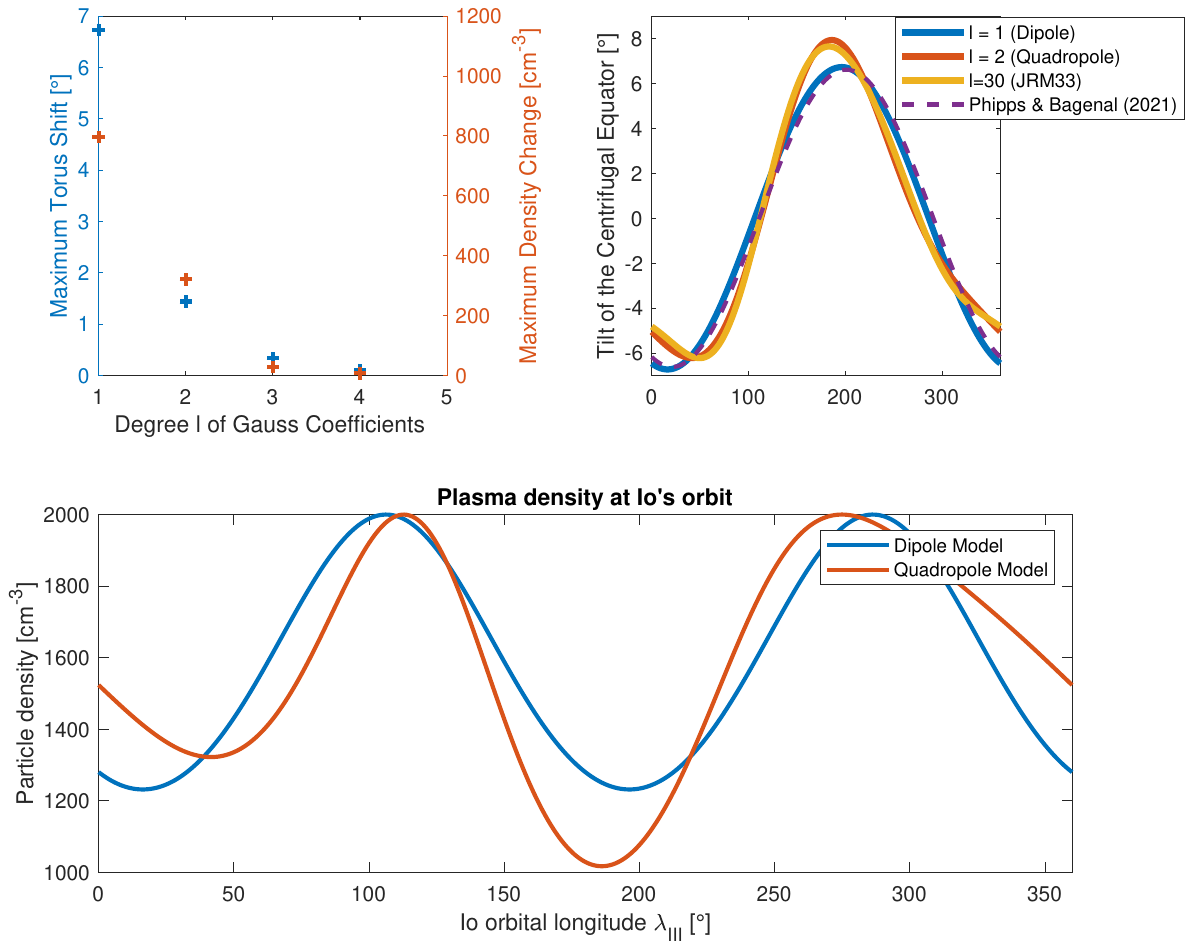}
\caption{ The position of the centrifugal equator has been calculated with different degrees $l$ of the JRM33 model. The position of the centrifugal equator relative to the rotational equator for dipole (l=1), quadropole (l=2) and full JRM33 model (l=30) is shown on the right upper panel and compared to the model by \citeA{phipps2021centrifugal} (Equation (2)) at the distance of Io as shown by the purple dashed line. From the variation for different degrees of Io's relative position to the torus center (blue in left panel), the maximum density variation due to the higher degrees in Io's vicinity has been calculated (red in left upper panel) using a scale height density model according to Equation (\ref{eq:densityModel}). As can be seen, the quadropole moment of the JRM33 model is sufficient to calculate the position of the centrifugal equator at Io's orbit. The lower panel shows the plasma number density at Io's orbit for the dipole and quadropole centrifugal equator model. A peak density of $n_0 = 2000$~cm$^{-3}$ and a scale height of $H = 1 R_J$ is used. The maximum difference between the two models is at $\lambda_{III} = 180^\circ$ at about $\Delta \rho \approx 250$~cm$^{-3}$.}
\label{fig:GaussTest}
\end{center}
\end{figure}

We conducted a study to investigate to what degree the JRM33 multipole moments influence the position of the Io plasma torus and therefore the density in Io's vicinity. In this study, we first calculated change in the position of the Io Plasma torus with each additional degree of the Gauss coefficients of the JRM33 model as can be seen in the upper left panel in Figure \ref{fig:GaussTest}. 
  From that, the variation of Io's relative position to the torus center due to each additional degree up to $l=5$ has been calculated (blue on upper right panel). We then used a torus density model according to Equation (\ref{eq:densityModel}) with a peak density of $\rho_0 = 2000$~cm$^{-3}$ and a scale height of $H = 1 R_J$ to calculate the maximum density change in Io's vicinity due to each additional degree. As can be seen, the density changes less then $\Delta\rho < 20$~cm$^{-3}$ for higher moments $l > 3$.  We therefore conclude that the quadropole moments are sufficient to describe the position of the torus. \\
  To estimate the effect of the shift in position of the plasma torus due to the quadropole moments on the plasma density in Io's vicinity, we calculated the density at Io's orbit for a dipole and quadropole model. The results are shown in the lower panel of Figure \ref{fig:GaussTest}. The largest discrepancy between the two models is around $\lambda_{III} = 180^\circ$, where the density differs about $\Delta \rho \approx 250$~cm$^{-3}$ or $\Delta \rho / \rho \approx 20\%$.

\subsubsection{Inversion of the Plasma Torus Position}
 
\begin{figure}
\begin{center}
\includegraphics[trim = 2cm 9.5cm 2cm 9.6cm, clip = true , width = 1\textwidth]{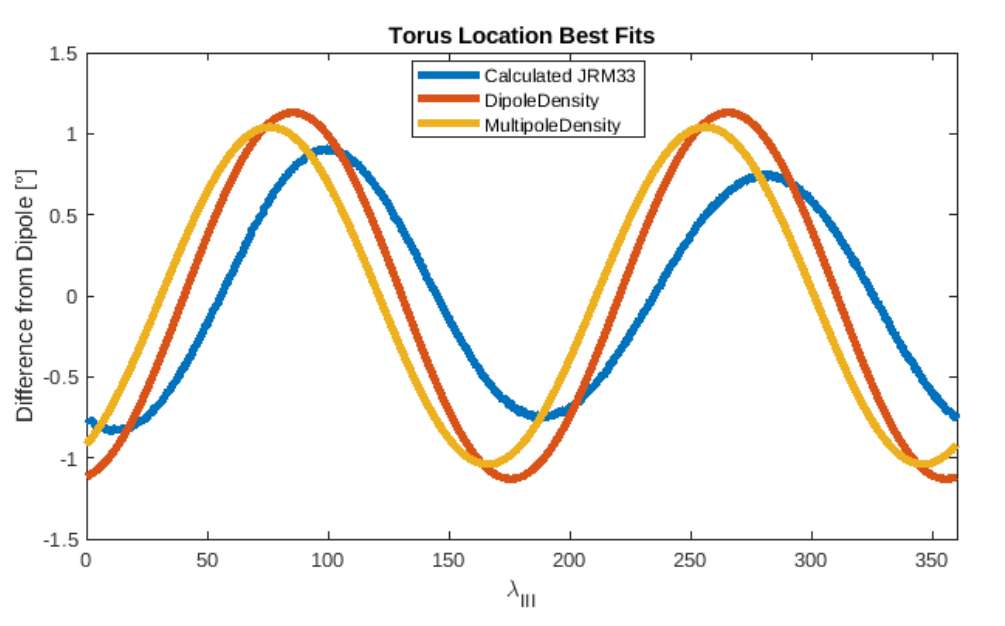}
\caption{Best fit models for the torus positions for the best fit peak density and scale height of the dipole (red) and multipole (yellow) model inversions. Phase and amplitude of both best fit models are comparable to the location of the JRM33 multipole centrifugal equator, shown in blue.  Therefore, the location of Io's footprint clearly indicates a pi-periodicity in the Alfvén wave travel times and therefore in Io's relative position to the torus center. A purely dipole centrifugal equator is not sufficient to explain the data.}
\label{fig:BestFitPosition}
\end{center}
\end{figure}

To test, whether the multipole centrifugal equator generally fits the data better, the position of the torus is also inverted. Since the data can be fitted by a large parameter space for two parameters already, we refrain from adding more inversion parameters. Instead, we use the values of peak density and scale height from the best fit models in the last section and use the amplitude $\theta_0$ and phase $\Delta \lambda$ of the pi-periodicity of the location of the torus corresponding to the quadropole moments as new inversion parameters. The lateral displacement $\theta$ of the torus to the rotational equator can be written as

\begin{equation}
    \theta(\lambda_{III}) = \theta_D(\lambda_{III}) + \theta_0 \sin(2 \lambda_{III} + \Delta \lambda),
\end{equation}

where $\theta_D$ is the tilt of the dipole centrifugal equator with $\theta_D(196.38^\circ) = -6.83^\circ$. 
The displacement from dipole centrifugal equator resulting from the inversions are shown in Figure \ref{fig:BestFitPosition}. The dipole model ( $n_0 = 1900$~cm$^{-3}$, $H = 1.01 R_J$) best fit parameters are $\theta_0 = 1.13^\circ$ and $\Delta \lambda = 81^\circ$ with a misfit of $\chi = 0.61$  compared to the previous misfit with $\theta_0 = 0^\circ$ of $\chi = 0.78$. The multipole model ($n_0 = 2133$~cm$^{-3}$ and $H = 1.07 R_J$) best fit parameters are $\theta_0 = 1.04^\circ$ and $\Delta \lambda = 62^\circ$ with a misfit of $\chi = 0.52$. Generally the fit improves, however not significantly for the multipole model, where the position of the torus already seems to be sufficient. The new best fit torus positions are comparable to the JRM33 multipole centrifugal equator position (blue line in Figure \ref{fig:BestFitPosition}) in phase and amplitude. This and the significant decrease in misfit for the dipole centrifugal equator model indicates that the torus is indeed located at the centrifugal equator of the JRM33 magnetic field model rather than a simple dipole centrifugal equator.

\section{Summary \& Conclusion}

We used Hubble Space Telescope observations of the Io Main Footprint as data to constrain a density model for the Io Plasma Torus. In this model we used the JRM33 magnetic field model by \citeA{connerney2022new} to map the magnetic field lines connecting the footprints to Io's orbit to calculate leading angle and Alfv\'en wave travel time. The travel time has then been used as data for a Monte-Carlo inversion to constrain peak density and scale height of the torus. In the first two inversions the position of the plasma torus is fixed once at the dipole centrifugal equator and once the multipole centrifugal equator of the JRM33 magnetic field model. The results show peak densities of $n_0 = (1900 \pm 321)$~cm$^{-3}$ and $n_0 = (2133 \pm 413)$~cm$^{-3}$ and scale heights of $H = (1.01 \pm 0.13) R_J$ and $H = (1.07 \pm 0.17) R_J$ for the dipole and multipole model, respectively. These values are in agreement, albeit generally lower than those of other models in the literature. Both models fit the data well. However, the misfit $\chi = 0.58$ of the multipole model is significantly lower than the misfit $\chi = 0.78$ of the dipole model. This agrees with a Monte Carlo test, where $91.4\%$ of the data points are better fitted by the multipole model. \\
In a second set of inversions the position of the plasma torus is fitted. The amplitude and phase shift of the lateral displacement is used as inversion parameter while scale height and peak density is kept fixed. The results show an agreement with the predicted JRM33 multipole centrifugal equator location of the Io Plasma Torus. \\
It could be shown that this method is suitable to constrain peak density and scale height of the Io Plasma Torus and yields results comparable to literature values.  We demonstrate quantitatively, that the torus is warped along the multipole centrifugal equator and the data can not sufficiently be explained by a simple dipole centrifugal equator. The latidudinal shift from a dipolar compared to a multipole centrifugal equator can differ by up to $1.5^\circ$ which translates to a change of Io's relative position to the torus center to up to $0.15 R_J \approx 6 R_{Io}$. In addition of the synodic period variation of $\Delta \rho \approx 800$~cm$^{-3}$, Io is exposed to a half synodic density variation of $\Delta\rho \approx 250$~cm$^{-3}$, which corresponds to a maximum in relative change of $\Delta \rho / \rho = 20\%$ . This needs to be included in high precision models of the Io plasma interaction to, for example, model the atmospheric sputtering processes or the evolution of the Io footprint brightness. The latter might be less faint near the minimum around 180 degrees compared to minimum around 330 degree \cite{wannawichian2010ten}. \\
The method presented here uses the integrated travel times of the Alfv\'en waves and is therefore able to constrain the mass density along the Io flux tube. However, the currently available data is not sufficient to distinguish between different species and scale heights of different populations. Furthermore, the non-uniqueness of the inversion method hinders an interpretation regarding a more complex density model. Nevertheless, with additional observations and more accurate positions of the Io main and reflected footprint this method could provide further insights into the density structure along the Io flux tube. Additional data could be used to constrain longitudinal and time variability and the density model could be adapted to incorporate the effect of different species and scale heights.

\section{Open Research}

 The processed travel times according to Equation (\ref{eq:travelTime}) and Figure \ref{fig:TravelTimeFit}, the used magnetic field mapping using the JRM33 model and the inversion results as shown in Figure \ref{fig:MonteCarlo} are available and published in \citeA{Schlegel2023a}.

\acknowledgments
 This project has received funding from the European Research Council (ERC) under the European Union’s Horizon 2020 research and innovation programme (grant agreement No. 884711).

\bibliography{bibfile}

\end{document}